\documentclass[12pt]{article}
\usepackage{graphicx} 
\begin{document}
\centerline{\bf Majority-vote on directed Small-World networks}

\bigskip
\centerline{Edina M.S. Luz$^1$ and F.W.S. Lima$^2$}
 
\bigskip
\noindent
$^1$Departamento de F\'{\i}sica, 
Universidade Estadual do Piau\'{\i}, 64002-150, Teresina - PI, Brazil\\
$^2$Departamento de F\'{\i}sica, 
Universidade Federal do Piau\'{\i}, 64049-550, Teresina - PI, Brazil

\medskip
  e-mail: edina@uespi.br, wel@ufpi.br
\bigskip
 
{\small Abstract: On {\it directed} Small-World networks the
 Majority-vote model with noise is now studied through Monte Carlo simulations.
In this model,  the order-disorder phase transition of the
 order parameter
 is well defined in this system. We calculate the value of the critical 
 noise parameter $q_{c}$ for several values of rewiring probability $p$  of the 
 directed Small-World network. The critical 
 exponentes $\beta/\nu$, $\gamma/\nu$ and
 $1/\nu$ were calculated for
 several values of $p$.}
 
 
\bigskip

 {\bf Introduction}

 Sumour and Shabat \cite{sumour,sumourss} investigated Ising models on 
 {\it directed} Barab\'asi-Albert networks \cite{ba} with the usual Glauber
 dynamics.  No spontaneous magnetisation was 
 found, in contrast to the case of {\it undirected}  Barab\'asi-Albert networks
 \cite{alex,indekeu,bianconi} where a spontaneous magnetisation was
 found lower a critical temperature which increases logarithmically with
 system size. Lima and Stauffer \cite{lima} simulated
 directed square, cubic and hypercubic lattices in two to five dimensions
 with heat bath dynamics in order to separate the network effects  form
 the effects of directedness. They also compared different spin flip
 algorithms, including cluster flips \cite{wang}, for
 Ising-Barab\'asi-Albert networks. They found a freezing-in of the 
 magnetisation similar to  \cite{sumour,sumourss}, following an Arrhenius
 law at least in low dimensions. This lack of a spontaneous magnetisation
 (in the usual sense)
 is consistent with the fact
 that if on a directed lattice a spin $S_j$ influences spin $S_i$, then
 spin $S_i$ in turn does not influence $S_j$, 
 and there may be no well-defined total energy. Thus, they show that for
 the same  scale-free networks, different algorithms give different
 results. It has been argued that nonequilibrium stochastic spin systems on 
 regular square lattice with up-down symmetry fall in the universality
 class of the equilibrium Ising model \cite{g}. This conjecture was
 found in several models that do not obey detailed balance \cite{C,J,M,mario,lima01}.
 Lima \cite{lima1,lima2} investigated the majority-vote model
 on {\it directed} and {\it undirected}
 Barab\'asi-Albert network and calculated the
 $\beta/\nu$, $\gamma/\nu$, and $1/\nu$ exponents and  these are 
 different from the Ising model and depend on the values of
 connectivity $z$ of the 
 Barab\'asi-Albert network.
 Campos $et$ $al$. \cite{campos} investigated the majority-vote model 
 on {\it undirected} small-world network by rewiring the two-dimensional square
 lattice. These
 small-world networks, aside from presenting quenched disorder, also 
 possess long-range interactions. They found that the critical exponents 
 $\gamma/\nu$ and $ \beta/\nu$ are different from the Ising model and depend
 on the rewiring probability. However, it was not evident whether the
 exponent change was due to the disordered nature of the network or due to 
 the presence of long-range interactions. Lima $et$ $al$. \cite{lima0} 
 studied the majority-vote model on Voronoi-Delaunay random lattices 
 with periodic boundary conditions. These lattices posses natural quenched
 disorder in their conecctions. They showed that presence of quenched 
 connectivity disorder is enough to alter the exponents $\beta/\nu$
 and $\gamma/\nu$ the pure model and therefore that is a relevant term to
 such non-equilibrium phase-transition. Now, we calculate the same
 $\beta/\nu$, $\gamma/\nu$, and $1/\nu$ exponents for
 majority-vote model on 
 {\it directed} small-world networks of S\'anchez et al. \cite{sanches}, 
 and for these networks the exponents are 
 different from the two-dimensional Ising model and independent on the values of
 rewiring probability $p$ of the 
 {\it directed} small-world networks. Here we study the cases for $p>0$, but we verify that when $p=0$ our results are agree with results of J. M. Oliveira \cite{mario}.
 
\bigskip
 
\begin{figure}[hbt]
\begin{center}
\includegraphics [angle=-90,scale=0.6]{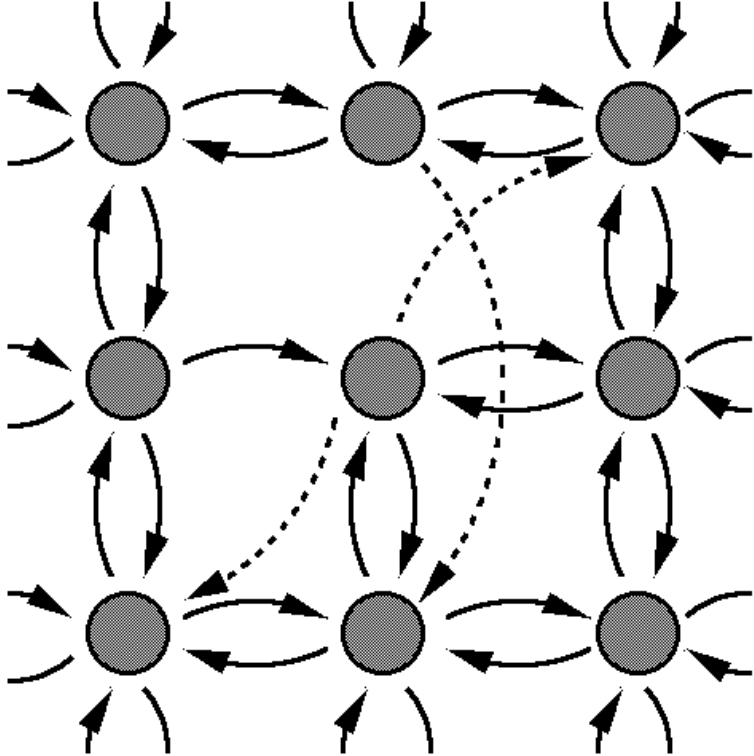}
\end{center}
\caption{Sketch of a {\it directed} small-world networks constructed
from a square regular lattice in $d=2$. Figure gentily yielded by Juan M. Lopez of the paper of S\'anchez et al. \cite{sanches}
 .}
\end{figure}

\bigskip

\begin{figure}[hbt]
\begin{center}
\includegraphics [angle=-90,scale=0.46]{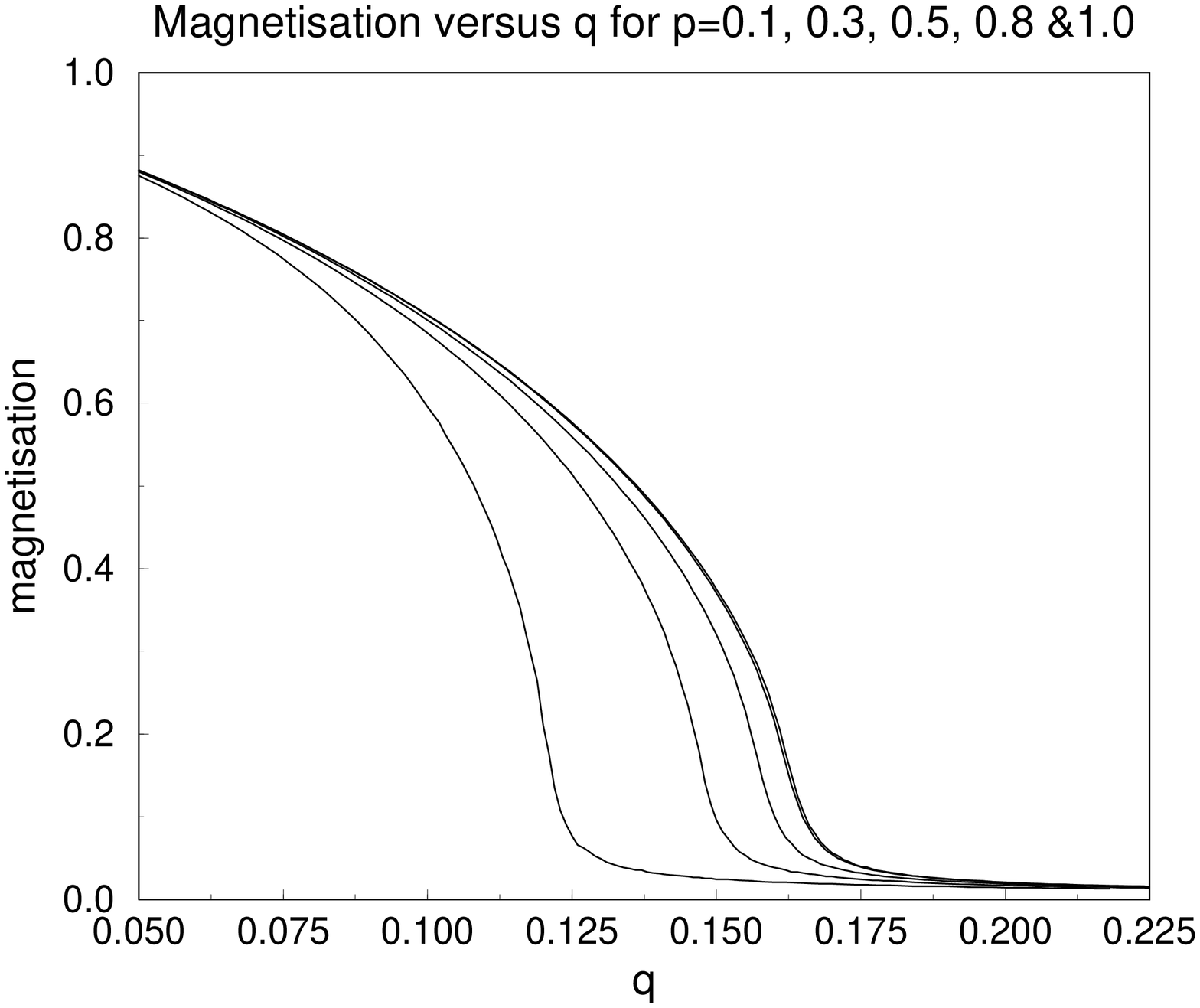}
\includegraphics [angle=-90,scale=0.46]{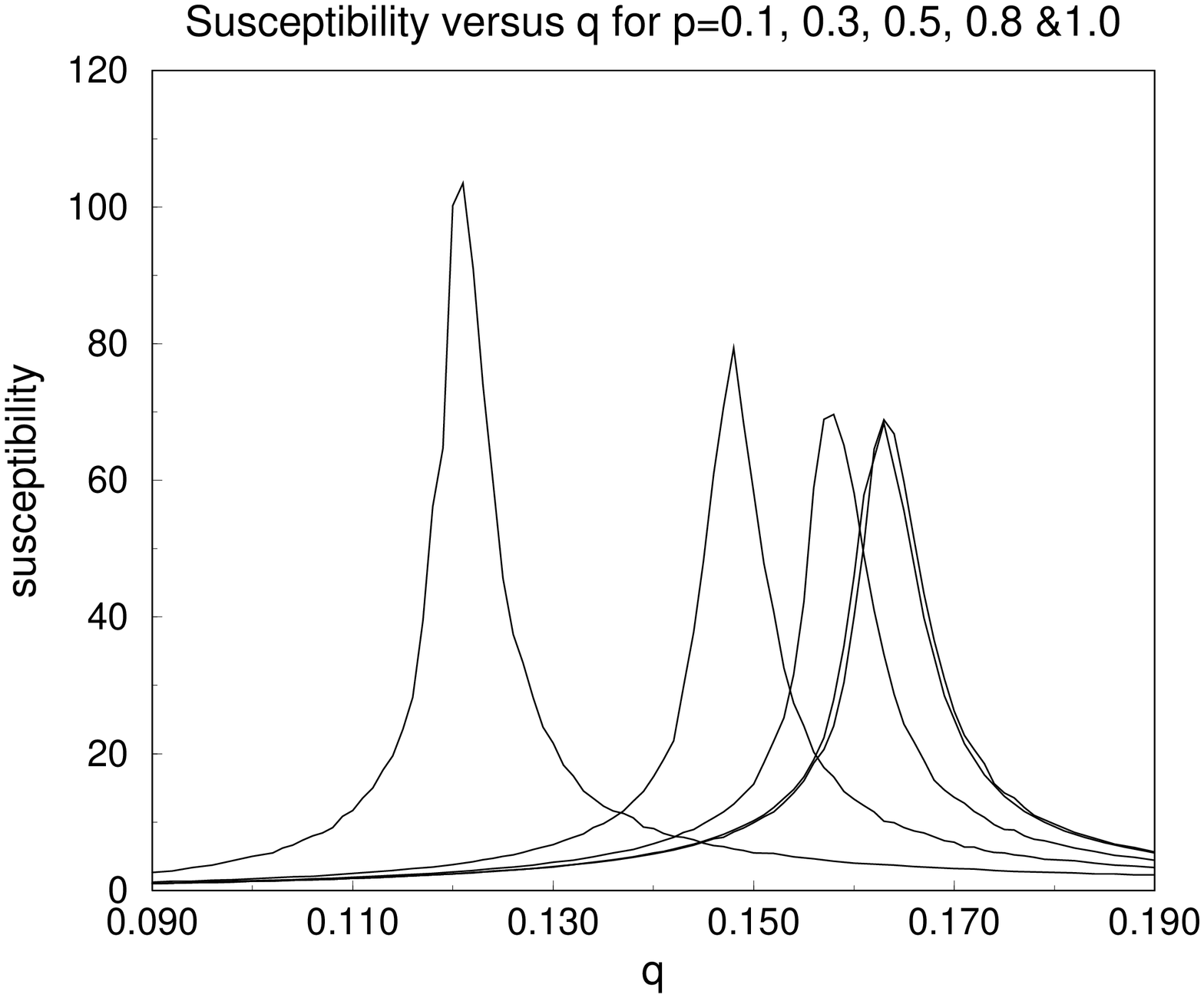}
\end{center}
\caption{
Magnetisation and susceptibility as a function of the noise parameter $q$, for
$N=16384$ sites. From left to right we have $p=0.1$, $0.3$, $0.5$, $0.8$,
and $1.0$ .}
\end{figure}
  
\bigskip
 
\begin{figure}[hbt]
\begin{center}
\includegraphics[angle=-90,scale=0.46]{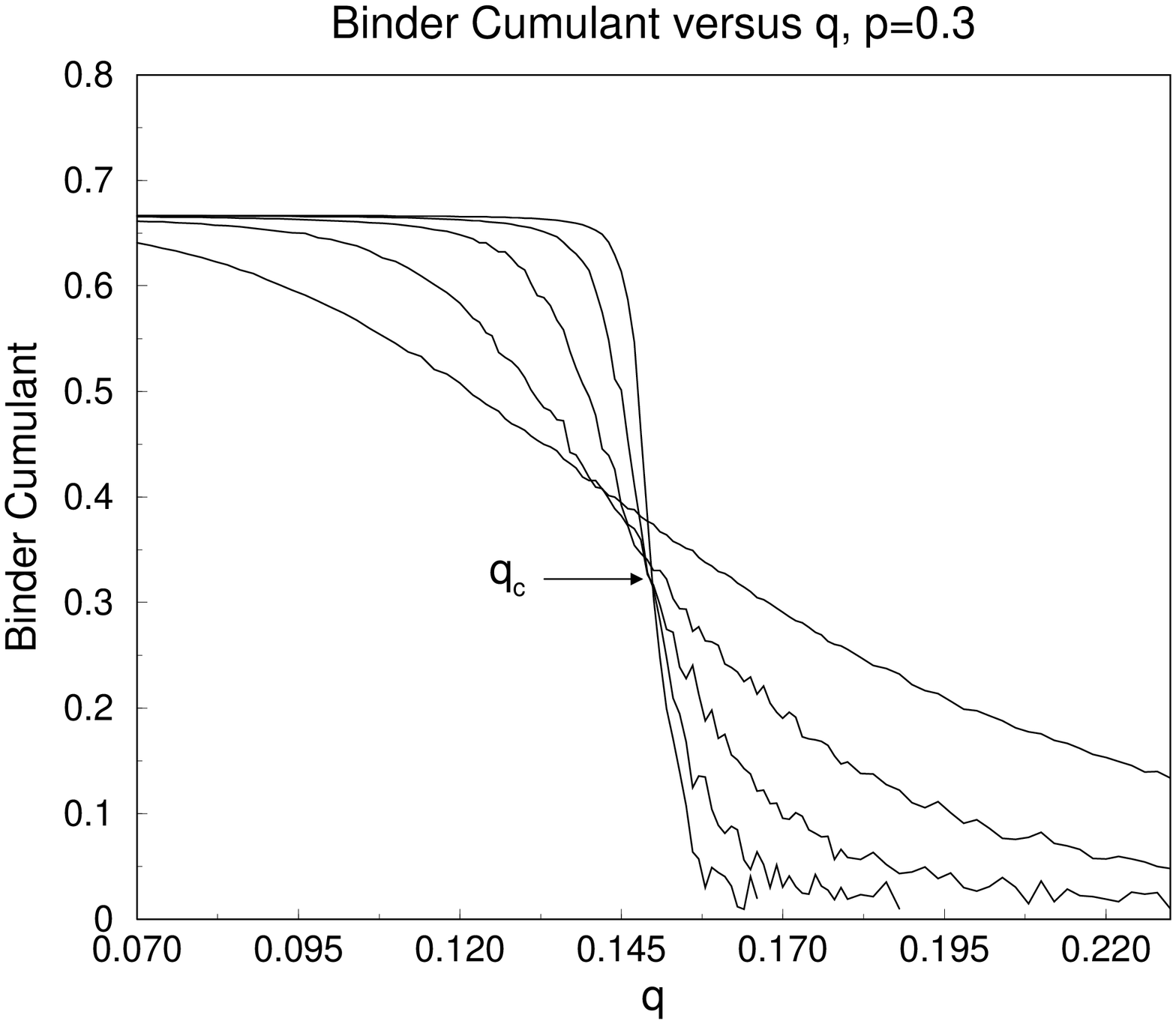}
\includegraphics[angle=-90,scale=0.46]{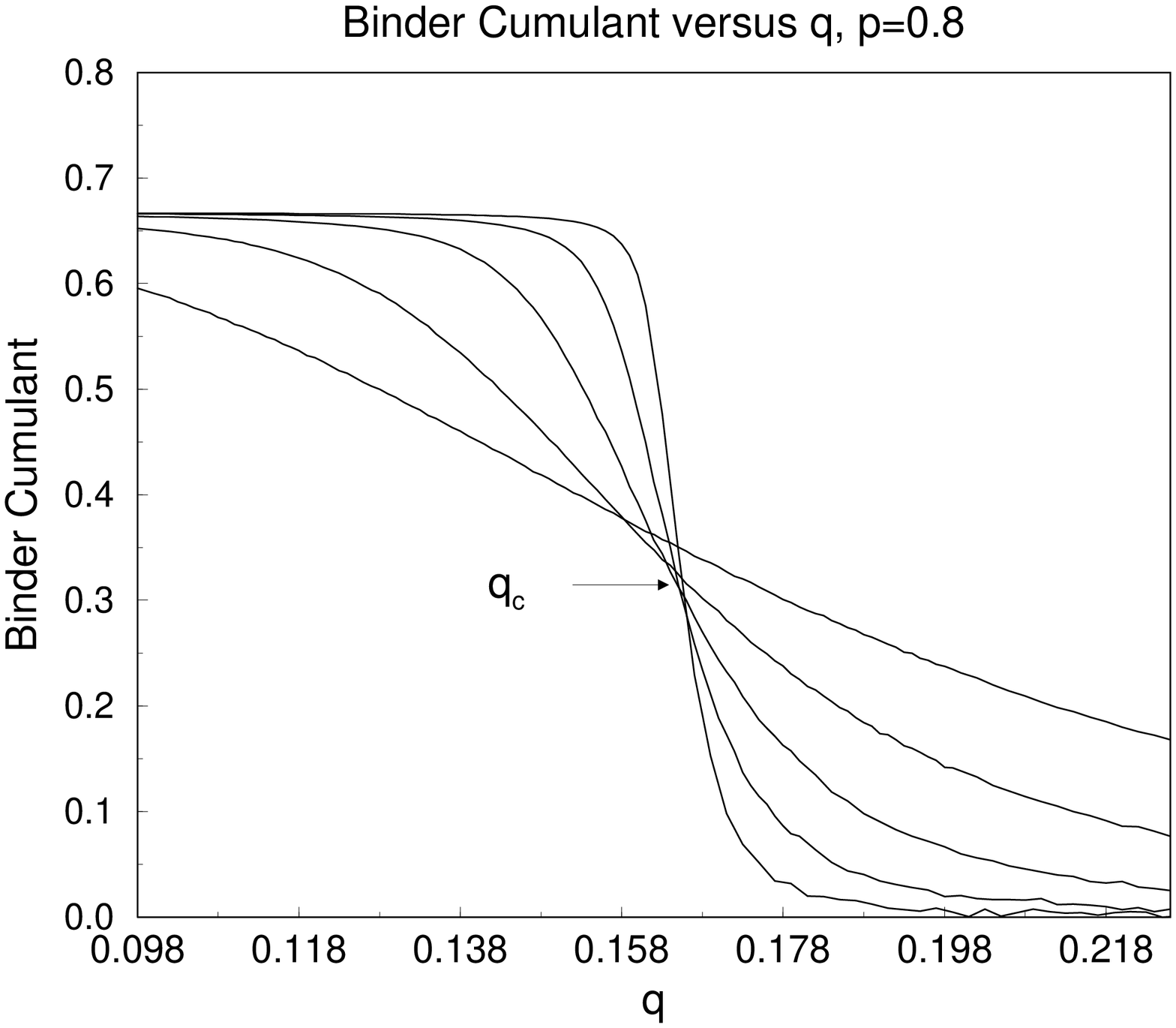}
\end{center}
\caption{
Binder's fourth-order cumulant as a function of $q$. We have $p=0.3$ and
 $p=0.8$ for $L=8$, $16$, $32$, $64$ and $128$.}
\end{figure}
 
\begin{figure}[hbt]
\begin{center}
\includegraphics[angle=-90,scale=0.60]{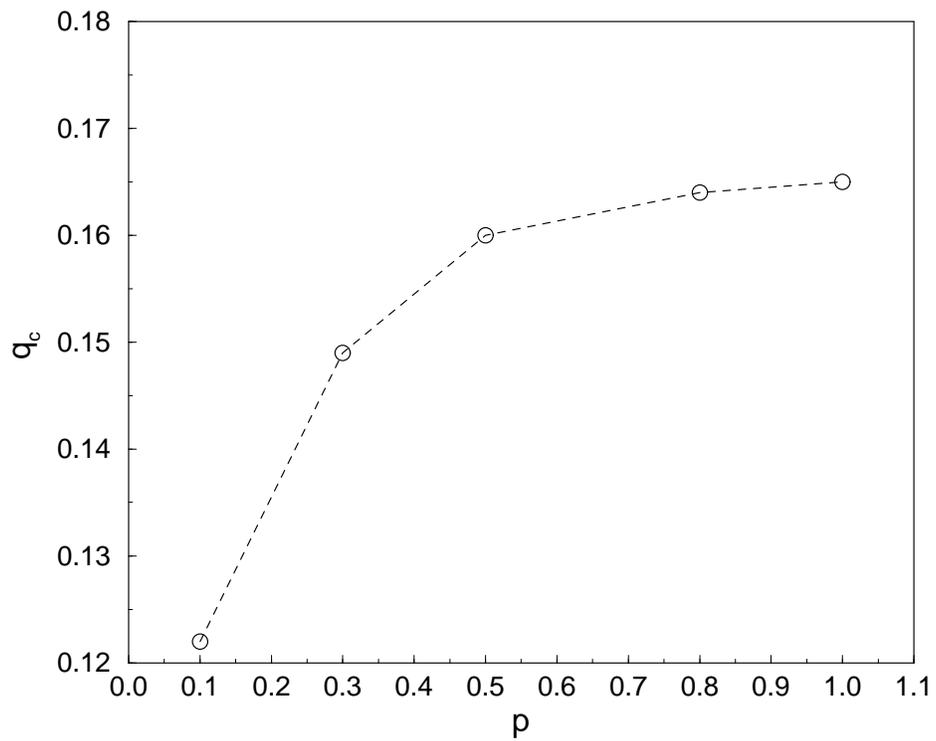}
\end{center}
\caption{The phase diagram, showing the dependence of critical 
noise parameter $q_{c}$ on probability $p$.
} 
\end{figure}
 
\begin{figure}[hbt]
\begin{center}
\includegraphics[angle=-90,scale=0.60]{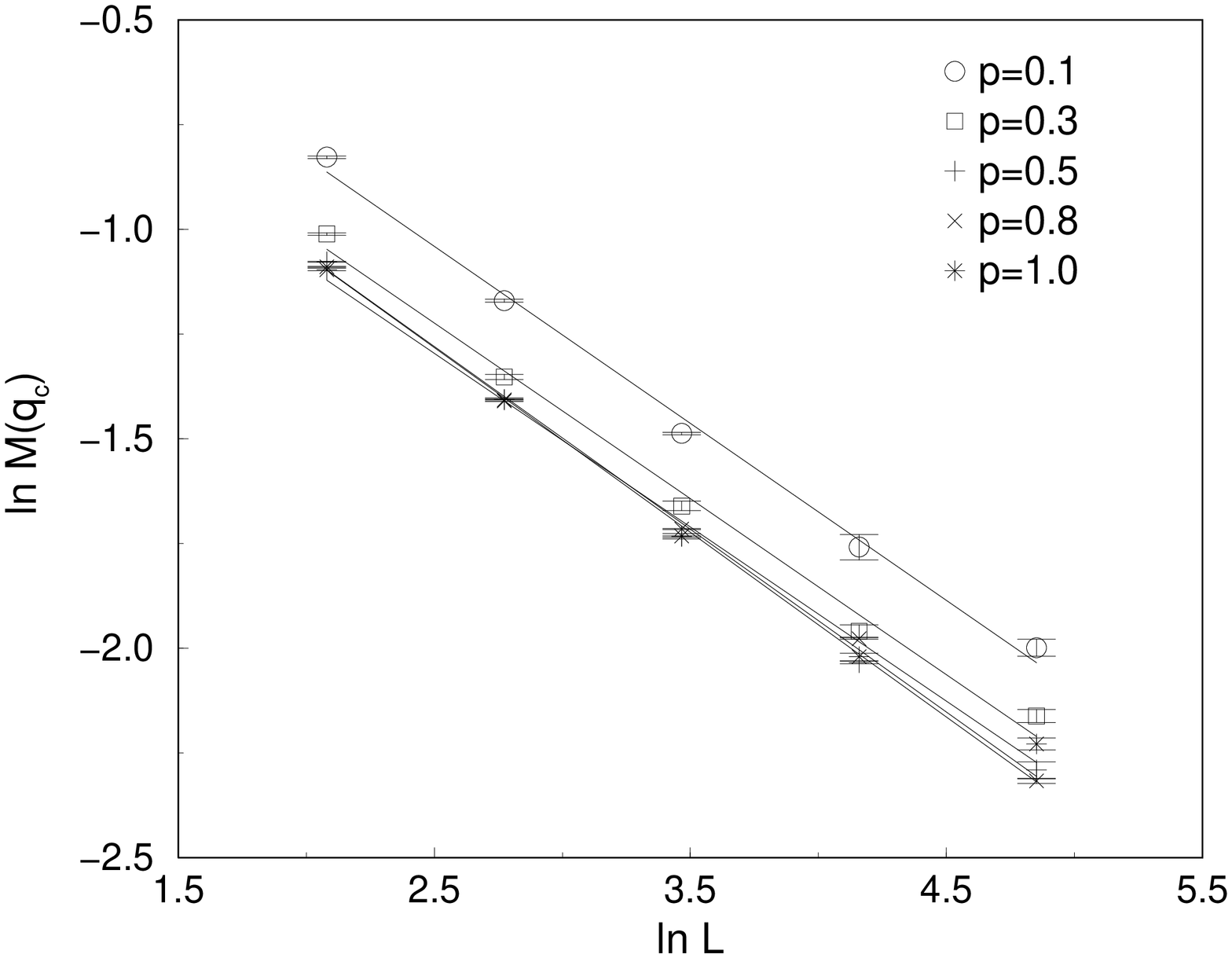}
\end{center}
\caption{ln $M(q_{c})$ versus  ln $L$. From top to bottom, $p=0.1$, $0.3$, $0.5$, $0.8$,
 and $1.0$. }
\end{figure}

\bigskip

\begin{figure}[hbt]
\begin{center}
\includegraphics[angle=-90,scale=0.60]{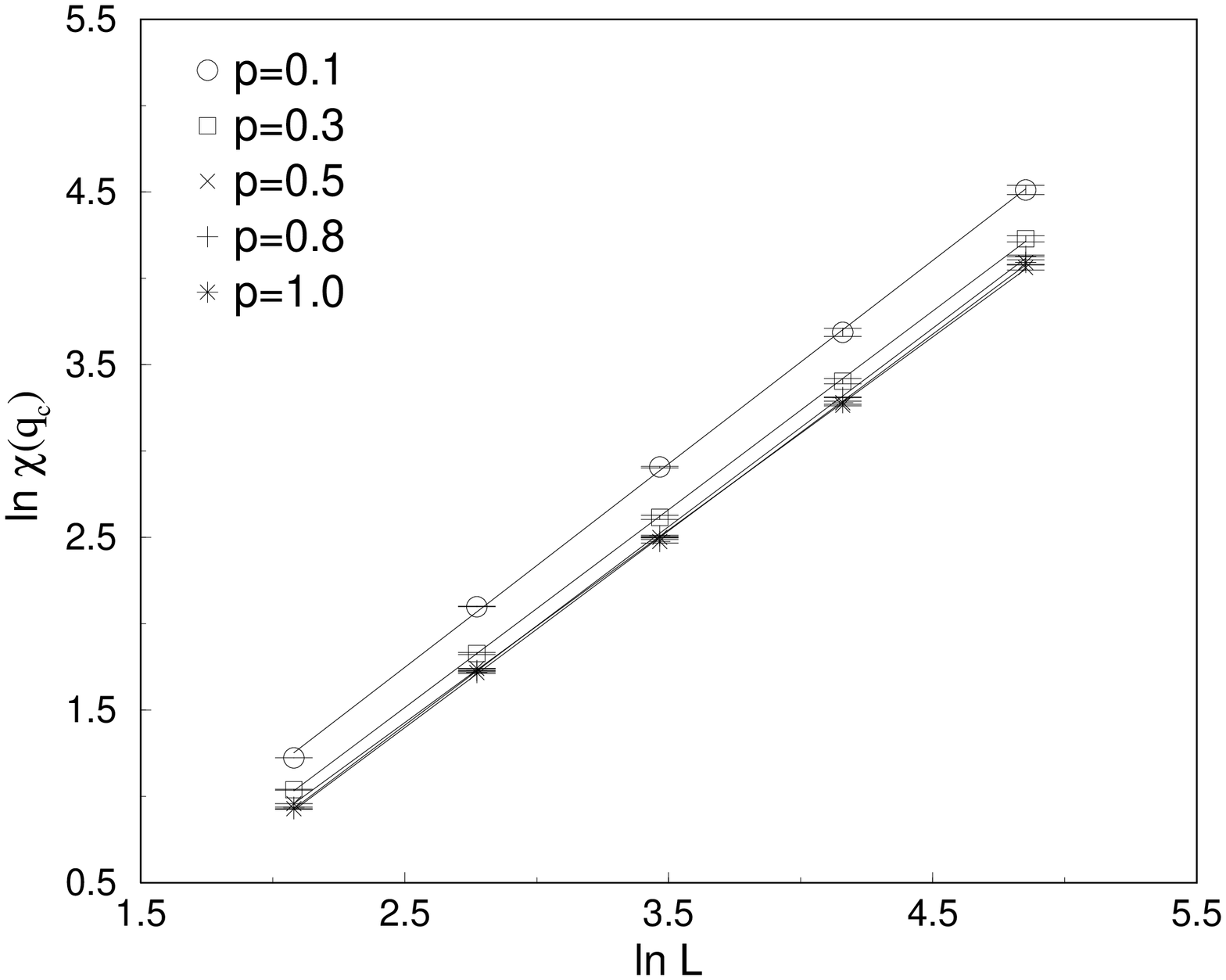}
\end{center}
\caption{ln $\chi(q_{c})$ versus  ln $L$. From top to bottom $p=0.1$, $0.3$, $0.5$, $0.8$,
 and $1.0$.
}
\end{figure}
\bigskip
 
\begin{figure}[hbt]
\begin{center}
\includegraphics[angle=-90,scale=0.60]{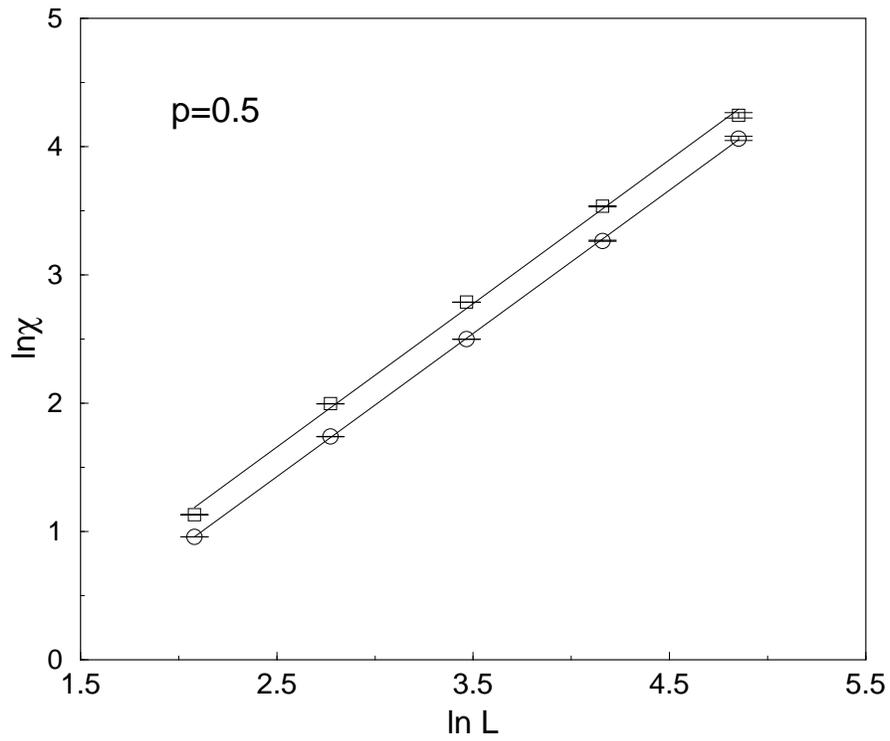}
\end{center}
\caption{ Plot of ln $\chi^{max}(L)$ (square) and ln$ \chi(q_{c})$ (circle) versus ln $L$ for 
connectivity $p=0.5$.}
\end{figure}

\begin{figure}[hbt]
\begin{center}
\includegraphics[angle=-90,scale=0.60]{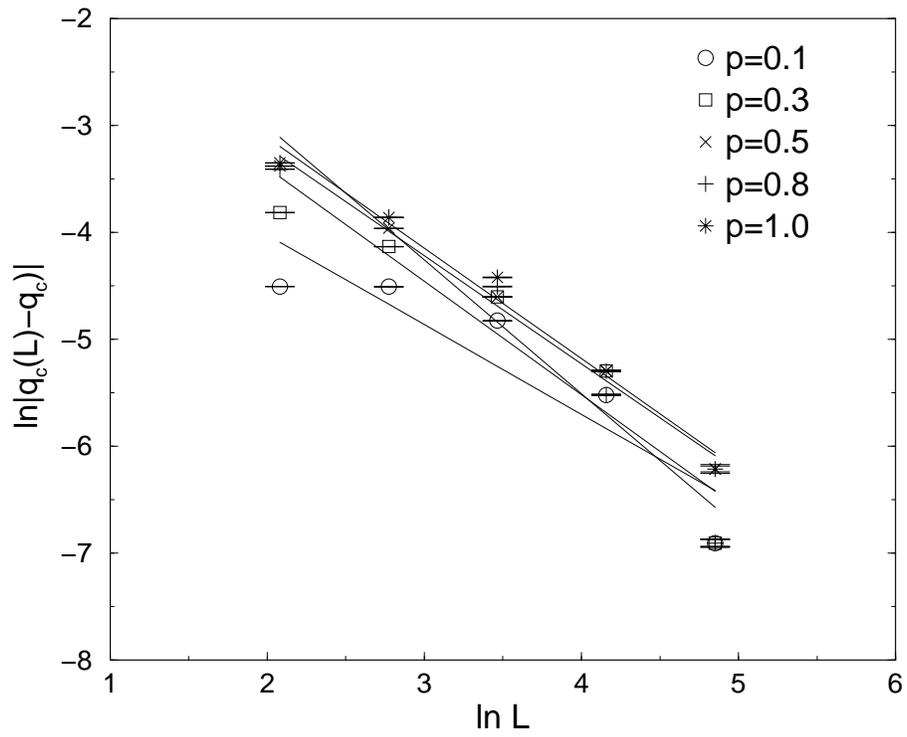}
\end{center}
\caption{ Plot of ln $|q_{c}(L)-q_{c}|$ versus ln $L$. From  bottom to top $p=0.1$, $0.3$, $0.5$, $0.8$,
 and $1.0$.}
\end{figure}

\bigskip

\begin{figure}[hbt]
\begin{center}
\includegraphics[angle=-90,scale=0.60]{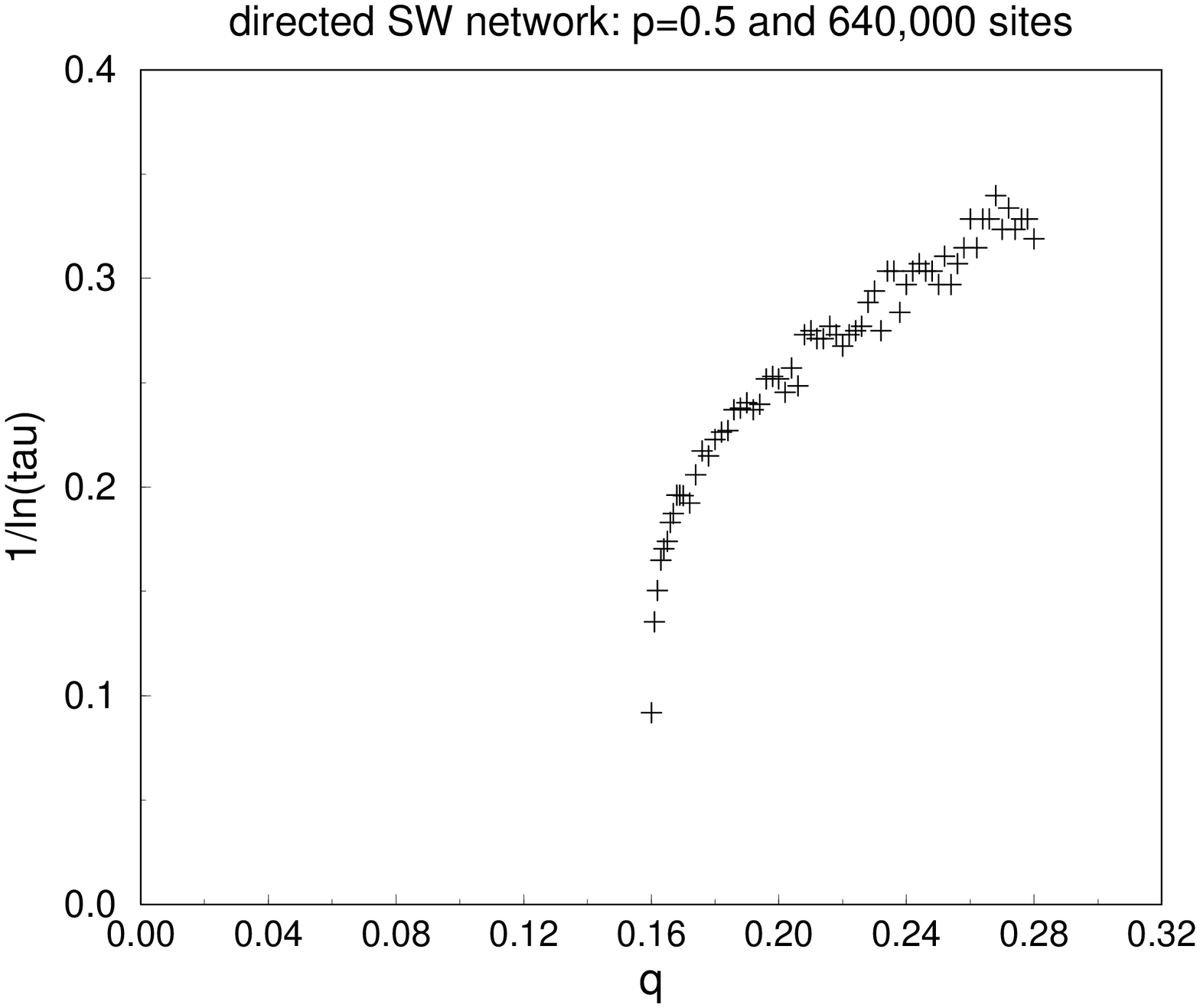}
\end{center}
\caption{ Plot of 1/ln(tau) versus $q$ for p=0.5.}
\end{figure}

\bigskip

{\bf Model and Simulaton}

We consider the majority-vote model, on {\it directed} 
small-world networks, defined \cite{mario,jff,lima0,pereira} by a set of
"voters" or spins variables ${\sigma}$ taking the values $+1$ or
$-1$, situated on every site of an {\it directed} 
small-world networks with $N=L \times L$ sites, were $L$ is the side of square lattice, and evolving in time by single spin-flip
like dynamics with a probability $w_{i}$ given by
\begin{equation}
w_{i}(\sigma)=\frac{1}{2}\biggl[ 1-(1-2q)\sigma_{i}S\biggl(\sum_{\delta
=1}^{k_{i}}\sigma_{i+\delta}\biggl)\biggl],
\end{equation}
where $S(x)$ is the sign $\pm 1$ of $x$ if $x\neq0$, $S(x)=0$ if $x=0$, and the 
sum runs over four  neighbours of $\sigma_{i}$. In this network, created for 
S\'anchez et al. \cite{sanches}, see Fig. 1, we start from a two-dimensional square
lattice consisting of sites linked to their four nearest neighbors by both outgoing and
incoming links. Then, with probability $p$, we reconnect nearest-neighbor outgoing
links to a different site chosen at random. After repeating this process for every
link, we are left with a network with a density $p$ of SW {\it directed} links. Therefore,
with this procedure every site will have exactly four outgoing links and
a varying (random) number of incoming links. The control parameter
$q$ plays the role of the temperature in equilibrium systems and measures
the probability of aligning antiparallel to the majority of neighbours.

To study the critical behavior of the model we define the variable
$m=\sum_{i=1}^{N}\sigma_{i}/N$. In particular, we were interested in the
magnetisation, susceptibility and the reduced fourth-order cumulant:
\begin{equation}
M(q)=[<|m|>]_{av},
\end{equation}
\begin{equation}
\chi(q)=N[<m^2>-<|m|>^{2}]_{av},
\end{equation}
\begin{equation}
U(q)=\biggl[1-\frac{<m^{4}>}{3<|m|>^{2}}\biggl]_{av},
\end{equation}
where $<...>$ stands for a thermodynamics average and $[...]_{av}$ square brackets
for a averages over the 20 realizations. 

These quantities are functions of the noise parameter $q$ and obey the finite-size
scaling relations

\begin{equation}
M=L^{-\beta/\nu}f_{m}(x)[1+ ...],
\end{equation}
\begin{equation}
\chi=L^{\gamma/\nu}f_{\chi}(x)[1+...],
\end{equation}
\begin{equation}
\frac{dU}{dq}=L^{1/\nu}f_{U}(x)[1+...],
\end{equation}
 where $\nu$, $\beta$, and $\gamma$ are the usual critical 
exponents, $f_{i}(x)$ are the finite size scaling functions with
\begin{equation}
x=(q-q_{c})L^{1/\nu}
\end{equation}
being the scaling variable, and the brackets $[1+...]$ indicate
corretions-to-scaling terms. Therefore, from the size dependence of $M$ and $\chi$
we obtained the exponents $\beta/\nu$ and $\gamma/\nu$, respectively.
The maximum value of susceptibility also scales as $L^{\gamma/\nu}$. Moreover, the
value of $q$ for which $\chi$ has a maximum, $ q_{c}^{\chi_{max}}=q_{c}(L)$,
is expected to scale with the system size as
\begin{equation}
q_{c}(L)=q_{c}+bL^{-1/\nu},
\end{equation}
were the constant $b$ is close to unity. Therefore, the  relations $(7)$ and $(9)$
are used to determine the exponent $1/\nu$.

We have performed Monte Carlo simulation on {\it directed} SW network with
various values of probability $p$. For a given $p$, we used systems
of size $L=8$, $16$, $32$, $64$, and $128$. We waited $10000$ Monte Carlo
steps (MCS) to make the system reach the steady state, and the time averages were
estimated from the next $ 10000$ MCS. In our simulations, one MCS is accomplished
after all the $N$ spins are updated. For all sets of parameters, we have generated
$20$ distinct networks, and have simulated $20$
independent runs for each distinct network.

\bigskip

{\bf Results and Discussion}

In Fig. 2 we show the dependence of the magnetisation $M$  and the susceptiblity
$\chi$  on the noise parameter, obtained from simulations on {\it directed}
SW network with $L=128 \times 128$ sites and several values of probability $p$.
In part (a) each  curve for $M$, for a given value of  $L$ and $p$, suggests
that there is  a phase transition from an ordered state to a disordered state. The
phase transition occurs at a value of the critical noise parameter $q_{c}$, which is
an increasing function the probability $p$  of the {\it directed}
SW network. In part (b) we
show the corresponding behavior of the susceptibility $\chi$, the value  of $q$
where  $\chi$ has a maximum is here identified as $q_{c}$. In Fig. 3  we plot
Binder's fourth-order cumulant for different values of $L$ and two different values
of $p$. The critical noise parameter $q_{c}$, for a given value of $p$, is estimated
as the point where the curves for different system sizes $L$ intercept each other.
In Fig 4 the phase diagram is shown as a function of the critical noise
parameter $q_{c}$ on probability $p$ obtained from the data of Fig. 3. 

The phase diagram of the majority-vote model on {\it directed}
SW network
shows that for a given network (fixed $p$ ) the system becomes ordered for
$q<q_{c}$, whereas it has zero magnetisation for $q\geq q_{c}$. We notice that the
increase of $q_{c}$ as a function the $p$ is slower that the one in \cite{pereira}. In
Figs. 5 and 6 we plot the dependence of the magnetisation and susceptibility, respectively, at $q=q_{c}$ versus the system
size $L$. The slopes of curves correspond to the exponent ratio $\beta/\nu$ and $\gamma/\nu$ of according
to Eq. (5) and (6), respectively.
The results show that the exponent ratio $\beta/\nu$ and $\gamma/\nu$ at $q_{c}$ are independent of $p$ (along with errors), see Table I.

In Fig. 7 we display the scalings for susceptibility at $q=q_{c}(L)$ (square), $\chi(q_{c}(L))$, and for its maximum amplitude, $\chi_{L}^{max}$, and the scalings for susceptibility at the $q=q_{c}$ obtained from Binder's cumulant, $ \chi(q_{c})$ (circle), versus $L$ for probability $p=0.5$. The exponents ratio $\gamma/\nu$ are obtained from the slopes
of the straight lines. For almost all the values of $p$, the exponents $\gamma/\nu$ of the two estimates agree (along with errors). We also observe that an increased $p$ does not mean a  tendency to increase or decrease the exponent ratio $\gamma/\nu$, see Table I. Therefore we can use the Eq. (9) , for fixed $p$, to obtain the critical exponent $1/\nu$, see Fig. 8. 

To improve our results obtained above we start with all spins up, a number of spins equal to $N=640000$, and time up 2,000,000 (in units of Monte Carlo steps per spins). Then we vary the noise parameter $q$ and at each $q$ study the time dependence for 9 samples. We determine the time $\tau$ after which the magnetisation has flipped its sign for fisrt time, and then take the median values of our nine samples. So we get different values $\tau_{1}$ for different noise parameters $q$. In Fig. 9  show that the decay time goes to infinity at some $q$ positive values this behavior sure that there is a phase transition for Majority-vote on
directed SW network.

The Table I summarizes the values of $q_{c}$, the exponents $\beta/\nu$, $\gamma/\nu$, and $1/\nu$ . J. M. Oliveira \cite{mario} showed that the majority-vote model  defined on a regular lattice has critical exponents that
fall into the same class of universality as the corresponding equilibrium Ising model. Campos et al \cite{campos} investigated the  critical behavior of the majority-vote on small-world networks by rewiring the two-dimensional square lattice, Pereira et al \cite{pereira} studied this model on Erd\"os-R\'enyi's random graphs, and Lima et al \cite{lima0} also studied this model on Voronoy-Delaunay
lattice and Lima on {\it directed} Barab\'asi-Albert network \cite{lima1}. The results obtained these authors show that the critical exponents of majority-vote model belong to different universality classes.

\begin{table}[h]
\begin{center}
\begin{tabular}{|c c c c c c|}
\hline
\hline
$ p $ & $q_{c}$ & $\beta/\nu$& ${\gamma/\nu}^{q_{c}}$ & ${\gamma/\nu}^{q_{c}(L)}$ & 
 $ 1/\nu$\\
\hline
$ 0.1 $ & $ 0.122(3) $ & $ 0.423(17) $ & $ 1.178(13) $ & $ 1.214(39) $ & $ 0.837(223)$\\

$ 0.3 $ & $ 0.149(3) $ & $ 0.419(21) $ & $ 1.148(5) $ & $ 1.152(28) $ & $ 1.059(208) $\\ 
$ 0.5 $ & $ 0.160(2) $ & $ 0.441(12) $ & $ 1.116(5) $ & $ 1.120(25) $ & $ 1.010(52) $\\
$ 0.8 $ & $ 0.164(2) $ & $ 0.436(9) $ & $ 1.149(5) $ & $ 1.117(23) $ & $ 1.248(158) $\\
$ 1.0 $ & $ 0.165(2) $ & $ 0.415(18) $ & $ 1.139(8) $ & $ 1.122(25) $ & $ 1.032(81) $\\

\hline
\hline
\end{tabular}
\end{center}
\caption{ The critical noise $q_{c}$, and the critical exponents
, for {\it directed} SW network with probability $p$. Error bars are statistical only.} \label{table1}
\end{table}

Finally, we remark that our MC results obtained on {\it directed} SW network for majority-vote model show that critical exponents are different from the results of 
\cite{mario} for regular lattice, of Pereira et al \cite{pereira} for Erd\"os-R\'enyi's random graphs, Lima \cite{lima1} and Campos $et$ $al$. \cite{campos} for
 on {\it undirected} SW network.

\bigskip
 
{\bf Conclusion}
 
In conclusion, we have presented a very simple nonequilibrium model on
{\it directed} SW network \cite{sanches}. In these networks, the majority-vote model presents a
second-order phase transition which occurs with 
probability $p\ge 0$. The exponents obtained are different from the other models \cite{mario, lima01, lima1, lima2, campos, pereira} suggesting that these exponents belong to another class of universality. 
However,  the exponents in the critical point $q_{c}$, $\beta/\nu$,
 $\gamma/\nu$ and  $1/\nu$ when $p$ grows no increase and do not decrease and are independents of a growing $p$ for $p>0$.

  The authors thank D. Stauffer for many suggestions and fruitful
discussions during the development this work and also for the revision of
this paper. We also acknowledge the Brazilian agency FAPEPI
(Teresina-Piau\'{\i}-Brasil) for  its financial support. This work also was supported the
system SGI Altix 1350 the computational park CENAPAD.UNICAMP-USP, SP-BRAZIL.

\end{document}